\documentclass[preprintnumbers,eqsecnum,prd,twocolumn,nofootinbib]{revtex4}  

\usepackage{graphicx}
\usepackage{latexsym}
\usepackage{amsmath}
\usepackage{amssymb}
\usepackage{amstext}

\def\beq{\begin{equation}}
\def\eeq{\end{equation}}

\begin{document}

\title{Detweiler's redshift invariant for spinning particles along circular orbits on a Schwarzschild background}

\author{Donato Bini$^1$, Thibault Damour$^2$, Andrea Geralico$^1$, Chris Kavanagh$^2$}
  \affiliation{
$^1$Istituto per le Applicazioni del Calcolo ``M. Picone,'' CNR, I-00185 Rome, Italy\\
$^2$Institut des Hautes \'Etudes Scientifiques, 91440 Bures-sur-Yvette , France.
}

\date{\today}

\begin{abstract}
We study the metric perturbations induced by a classical spinning particle moving along a circular orbit on a Schwarzschild background, limiting the analysis to effects which are first order in spin.
The particle is assumed to move on the equatorial plane and has its spin aligned with the $z$-axis.
The metric perturbations are obtained by using two different approaches, i.e., by working in two different gauges: the Regge-Wheeler gauge (using the Regge-Wheeler-Zerilli formalism) and a radiation gauge (using the Teukolsky formalism). 
We then compute the linear-in-spin contribution to the first-order self-force contribution to Detweiler's redshift invariant up to the 8.5 post-Newtonian order. We check that our result is the same in both gauges, as appropriate for a gauge-invariant quantity, and agrees
with the currently known 3.5 post-Newtonian results.
\end{abstract}


\maketitle

\section{Introduction}

In the new field of gravitational wave astrophysics, an interesting potential source are extreme mass ratio inspirals, where a small compact body of mass $\mu$ orbits and eventually coalesces with a much more massive black hole of mass $M$, where $\mu/M\sim10^{-6}$. These systems are most commonly modelled using the gravitational self-force (GSF) approach. In this approach, in order to accurately model the inspiral waveform, one needs to account correctly for both dissipation of the orbital parameters and conservative shifts, which grow secularly when taken in conjunction with the dissipation. A significant focus of conservative GSF calculations has been on gauge-invariant, physical effects localized on the small mass $\mu$. These were initiated by Detweiler \cite{Detweiler:2008ft} who defined, and computed, a redshift variable for a particle on a circular orbit in Schwarzschild spacetime (i.e., the linear-in-mass-ratio contribution to $u^t$, the time component of the particle's 4-velocity). This provided the first identified conservative, gauge-invariant GSF effect (though it was not, initially, related to the dynamics of small-mass-ratio systems). Soon after, the GSF computation of
 shifts in the innermost stable circular orbit, and of precession of the periapsis \cite{Barack:2009ey,Barack:2010ny}
 provided other conservative, gauge-invariant GSF effects (of more direct dynamical significance).
 
 Detweiler's redshift computations were pushed to high numerical accuracy, and compared to the third
 post-Newtonian (3PN) analytical knowledge of comparable-mass binary systems \cite{Blanchet:2009sd,Blanchet:2010zd}.
 Moreover, the later discovery of the ``First Law of Binary Black Hole Mechanics" \cite{Tiec:2011ab}, allowed one to
 extract the dynamical significance of GSF redshift computations \cite{Tiec:2011dp,Barausse:2011dq}.
The first complete\footnote{For earlier computations of the logarithmic 4PN-level contribution, see Refs. \cite{Damour:2009sm} and \cite{Blanchet:2010zd}.} analytic self-force computation of Detweiler's redshift invariant at the fourth post-Newtonian (4PN) was performed  by Bini and Damour \cite{Bini:2013zaa}, who showed how to combine the Regge-Wheeler-Zerilli \cite{Regge:1957td,Zerilli:1971wd} (RWZ) formalism for the Schwarzschild gravitational perturbations with the hypergeometric-expansion analytical solutions of the RWZ radial equation obtained by Mano, Suzuki and Takasugi \cite{Mano:1996mf,Mano:1996vt} (MST). The methodology given in Ref.  \cite{Bini:2013zaa} allowed the extension to higher PN levels: indeed, results were soon derived at the 6PN level \cite{Bini:2013rfa},  the 8.5PN one \cite{Bini:2014nfa}, the 9.5PN one \cite{Bini:2015bla}, ending, with a considerable jump, at the 22.5PN one \cite{Kavanagh:2015lva}.

In the meantime the GSF community became interested in computing other gauge invariant quantities, associated with spin precession along circular orbits in Schwarzschild \cite{Dolan:2013roa,Bini:2014ica,Bini:2015mza} and tidal invariants (quadrupolar, octupolar) again along circular orbits in Schwarzschild \cite{Dolan:2014pja,Bini:2014zxa,Nolan:2015vpa}; while most of these works contained strong field numerics or analytic PN calculations, other conceptually useful methods were also introduced (e.g., the PSLQ reconstruction of fractions, see Ref. \cite{pslq}).
	
In addition to defining new invariants, considerable work has been ongoing in extending GSF computations towards more astrophysically relevant scenarios. For example, the redshift and spin precession invariants along eccentric (equatorial) orbits in Schwarzschild have been studied \cite{Barack:2011ed,Akcay:2015pza,Tiec:2015cxa,Bini:2015bfb,Hopper:2015icj,Bini:2016qtx,Akcay:2016dku,Kavanagh:2017wot}.
Including for the first time spin on the primary black hole, Abhay Shah gave in 2015 the first (4PN) GSF computation of the redshift invariant along circular orbits in Kerr spacetime \cite{shah_MG14,Johnson-McDaniel:2015vva}. This PN calculation was then extended in Refs. \cite{Bini:2015xua,Kavanagh:2016idg} and calculated for eccentric orbits in Ref. \cite{Bini:2016dvs}. While formulations have been provided for spin precession in Kerr spacetime \cite{Akcay:2017azq}, the practical calculation of further gauge invariants or the generalisation to inclined orbits have been halted by technical difficulties in the regularization procedures and metric completion of the non-radiative multipoles. However, significant recent work, including numerical calculations of the full self-force for generic inclined eccentric orbits in Kerr, show that these issues are in principle solved \cite{Merlin:2016boc,vandeMeent:2016hel,vandeMeent:2017fqk,vandeMeent:2017bcc}.

One of the strong motivations for the analytic GSF-PN computational effort has been the possibility to convert such high PN-order GSF information into other approximation formalisms useful for computing (comparable-mass) binary inspirals, such as the Effective-One-Body (EOB) model \cite{Buonanno:1998gg,Buonanno:2000ef,Damour:2001tu}.
For example, Damour \cite{Damour:2009sm} showed how to compute some combinations of EOB radial potentials 
from GSF data. Further
 use of the first law of mechanics \cite{Tiec:2011ab} allowed the computation of individual EOB radial potentials \cite{Barausse:2011dq,Akcay:2012ea}.  Following this, high-order PN computations of these potentials were actually accomplished \cite{Bini:2015xua,Bini:2015bfb,Bini:2016qtx,Bini:2016dvs}. It  has been shown that the knowledge of the eccentric redshift invariant maps completely the non-spinning effective-one-body Hamiltonian \cite{Tiec:2015cxa}. Transcription of information from GSF to the spinning EOB Hamiltonian remains ongoing.
 
The aim of this paper is to provide a generalisation of Detweiler's redshift in Schwarzschild spacetime to the case where the small body $\mu$ has a small but non-negligible spin $s$, and to provide an 8.5PN-accurate post-Newtonian expansion valid to linear order in both the mass-ratio and the spin. Test spinning particles no longer move on geodesics of the spacetime but experience a force due to the coupling of the spin of the body and the Riemann curvature tensor of the background which must be included, according to the Mathisson-Papapetrou-Dixon (MPD) model \cite{Mathisson:1937zz,Papapetrou:1951pa,Dixon:1970zza}. Hence, we shall consider a particle moving along an accelerated circular orbit. Metric perturbations
generated by spinning particles  (both in Schwarzschild and in Kerr spacetimes) have been considered, e.g., in  Refs. \cite{Mino:1995fm,Tanaka:1996ht,Saijo:1998mn} (see also the review by Sasaki and Tagoshi \cite{Sasaki:2003xr}), with
the aim of computing  the emitted fluxes of gravitational wave. Similarly, PN calculations involving spinning bodies also exist \cite{Tagoshi:2000zg,Faye:2006gx}. 

To our knowledge our study is the first analytic calculation of a conservative effect of the self-force for a spinning particle. To internally validate our results we perform all calculations both in the Regge-Wheeler (RW) gauge (solving the RWZ equations) and in the (outgoing) radiation gauge (using the Teukolsky approach). The exception to this is the low-multipole problem, for which we use the RWZ approach in both cases. As an important side result of our work, we explicitly give the complete (interior and exterior) metric perturbations for $\ell=0,1$, which are needed for any other study of conservative effects of the self-force. An independent check of the first few PN orders in our results for the redshift are given by comparing with the corresponding comparable-mass redshift, derived from currently known PN results \cite{Blanchet:2012at,Levi:2015uxa}. This yields complete agreement, thereby getting strong validation for our formulation and methods.

We follow the notation and convention of previous GSF papers, e.g., \cite{Bini:2013rfa}. [Note that we use
here the notation $\mu$ (instead of $m_1$) for the small mass, and $M$ (instead of $m_2$) for the large mass.]
The metric signature is chosen to be $-+++$ and units are such that $c=G=1$ unless differently specified.
Greek indices run from 0 to 3, whereas Latin ones from 1 to 3.

\section{Spinning particle motion in the background Schwarzschild spacetime}

Our background  Schwarzschild spacetime has a line element, written in standard coordinates $(t,r,\theta,\phi)$,  given by
\begin{eqnarray} 
\label{metric}
d  s^2 &=&\bar g_{\alpha\beta}dx^\alpha dx^\beta \\
&=& -fd t^2 + f^{-1} d r^2 
+ r^2 (d \theta^2 +\sin^2 \theta d \phi^2)\,,\nonumber
\end{eqnarray}
where $f=1-2M/r$. Let us first introduce an orthonormal frame adapted to the static observers, namely those at rest with respect to the space coordinates
\begin{eqnarray}
\label{frame}
e_{\hat t}&=&f^{-1/2}\partial_t\,, \quad
e_{\hat r}=f^{1/2}\partial_r\,, \nonumber\\
e_{\hat \theta}&=&\frac{1}{r}\partial_\theta\,, \quad
e_{\hat \phi}=\frac{1}{r\sin \theta}\partial_\phi\,,
\end{eqnarray}
where $\{\partial_\alpha\}$ is the coordinate frame.
As a convention, the physical (orthonormal) component along $-\partial_\theta$ which is perpendicular to the equatorial plane will be referred to as ``along the positive $z$-axis" and will be indicated by the index $\hat z$, when convenient: $e_{\hat z}=-e_{\hat \theta}$. Furthermore, we indicate with a bar background quantities to be distinguished from corresponding perturbed spacetime quantities. 

The Mathisson-Papapetrou-Dixon (MPD) equations \cite{Mathisson:1937zz,Papapetrou:1951pa,Dixon:1970zza} governing the motion of a spinning test particle in a given gravitational background read
\begin{eqnarray}
\label{papcoreqs1}
\frac{{\rm D}\bar P^{\mu}}{d \bar\tau} & = &
- \frac12 \, R^\mu{}_{\nu \alpha \beta} \, \bar U^\nu \, S^{\alpha \beta}
\,,
\\
\label{papcoreqs2}
\frac{{\rm D}S^{\mu\nu}}{d \bar\tau} & = & 
2 \, \bar P^{[\mu}\bar U^{\nu]}
\,,
\end{eqnarray}
where $\bar P^{\mu} \equiv \mu \bar u^\mu$ (with $\bar u \cdot \bar u = -1$) is the total 4-momentum of the particle with mass $\mu$, $S^{\mu \nu}$ is a (antisymmetric) spin tensor, and $\bar U^\mu=d z^\mu/d\bar\tau$ is the timelike unit tangent vector of the \lq\lq center of mass line'' (with parametric equations $x^\mu=z^\mu(\bar\tau)$) used to make the multipole reduction, parametrized by the proper time $\bar \tau$. In order for the model to be mathematically self-consistent certain additional conditions should be imposed. As is standard, we adopt here the Tulczyjew-Dixon conditions~\cite{Dixon:1970zza,tulc59}, i.e.,
\beq
\label{tulczconds}
S^{\mu\nu} \bar P_{\nu} = \mu \, S^{\mu\nu}\bar u{}_\nu=0\,.
\eeq
Consequently, the spin tensor can be fully represented by a spatial vector (with respect to $\bar u$),
\beq
\label{spinvecdef}
S(\bar u)^\alpha=\frac12 \eta(\bar u)^\alpha{}_{\beta\gamma}S^{\beta\gamma}
\,,
\eeq
where $\eta(\bar u)_{\alpha\beta\gamma}=\eta_{\mu\alpha\beta\gamma}\bar u^\mu $ 
is the spatial unit volume 3-form (with respect to $\bar u$) built from the unit volume 4-form
$\eta_{\alpha\beta\gamma\delta}=\sqrt{-\bar g}\, \epsilon_{\alpha\beta\gamma\delta}$,
with $\epsilon_{\alpha\beta\gamma\delta}$ ($\epsilon_{0123}=1$) being the
Levi-Civita alternating symbol and $\bar g$ the determinant of the metric.

Both the mass $\mu \equiv (- \bar P_{\alpha} \bar P^{\alpha})^\frac12$, and the
the magnitude $s$ of the spin vector
\beq
\label{sinv}
s^2=S(\bar u)^\beta S(\bar u)_\beta = \frac12 S_{\mu\nu}S^{\mu\nu}\,, 
\eeq
are constant along the trajectory of a spinning particle, as  follows from Eqs. \eqref{papcoreqs1}, \eqref{papcoreqs2}, 
when using Eq. \eqref{tulczconds}. We shall endow here the spin magnitude $s$ with a positive (negative) sign
if its orbital angular momentum is parallel (respectively, antiparallel) to $e_{\hat z}=-e_{\hat \theta}$.
A requirement which is essential for the validity of the Mathisson-Papapetrou-Dixon model (and of the test particle approach) is that the characteristic length scale $|s|/\mu$ associated with the particle's internal structure be small compared to the natural length scale $M$ associated with the background field. 
Hence the following condition must be assumed: $|\hat s|\equiv|s|/(\mu M) \ll 1$. 
This leads one to consider only the terms of first order in the spin in Eqs. (\ref{papcoreqs1}) and (\ref{papcoreqs2}) and to neglect higher order terms.
As a result, the 4-momentum $\bar P$ is parallel to $\bar U$ to first order in $\hat s$, i.e., $\bar P= \mu \bar U+O(\hat s^2)$, and the spin tensor is parallel-transported along the path (from Eq. (\ref{papcoreqs2})).
 In particular under these assumptions we can  identify $\bar U^\mu \equiv \bar u^\mu$.

Finally, when the background spacetime has Killing vectors, there are conserved quantities along the motion \cite{ehlers77}. For example, in the case of stationary axisymmetric spacetimes with coordinates adapted to the spacetime symmetries,  $\xi=\partial_t$ is the timelike Killing vector and $\eta=\partial_\phi$ is the azimuthal Killing vector. The corresponding conserved quantities are the  energy $\bar E$ and the angular momentum $\bar J$ of the particle, namely
\begin{eqnarray}
\label{totalenergy}
\bar E&=&-\xi_\alpha \bar P^\alpha +\frac12 S^{\alpha\beta}\nabla_\beta \xi_\alpha \,, \nonumber\\
\bar J&=&\eta_\alpha \bar P^\alpha -\frac12 S^{\alpha\beta}\nabla_\beta \eta_\alpha \,,
\end{eqnarray}
where  $\nabla_\beta \xi_\alpha=-\frac{M}{r^2}\,\delta^{tr}_{\alpha\beta}$ and $\nabla_\beta \eta_\alpha=r\sin^2\theta\,\delta^{\phi r}_{\alpha\beta} 
$.

\subsection{Solution for a spinning test particle in circular motion in the Schwarzschild spacetime}

The MPD equations admit (to linear order in $\hat s$) the following solution for a spinning test particle moving along a circular orbit on the equatorial plane with spin vector $S(\bar U)=S^{\hat\theta}e_{\hat \theta}=s\,e_{\hat z}$ orthogonal to it  (see, e.g., Ref. \cite{Bini:2014poa}): 
\beq
\label{casocircU}
\bar U=\bar u^t(\partial_t+\Omega\partial_\phi)\,,
\eeq
with normalization factor
\begin{eqnarray}
\bar u^t&=&\frac{1}{\sqrt{1-3u}} \left(1- \frac32{\hat s}  \frac{u^{5/2}}{1-3u}\right)\,,
\end{eqnarray}
and angular velocity  
\beq
\label{zetaspin}
M\Omega=u^{3/2}\left(1-\frac32{\hat s}u^{3/2}\right)\,,
\eeq
where $u=M/r$ is the dimensionless inverse radial distance and ${\hat s}=s/(\mu M)$ is the dimensionless spin parameter introduced above.
A spatial triad adapted to $\bar U$ can be built with
\beq
E_1=e_{\hat r}\,,\quad E_2=e_{\hat \theta}\,,\quad E_3=\frac{r\Omega}{f^{1/2}} \bar u^t \left( \partial_t +\frac{f}{r^2\Omega} \partial_\phi \right)\,.
\eeq
These will be useful below in the definition of the stress tensor.

A key component of defining gauge invariant functions is to consider gauge-invariant quantities as functions of gauge invariant arguments. We shall use as gauge-invariant argument (to parametrize circular orbits) the dimensionless frequency variable $y=(M\Omega)^{2/3}$, so that from Eq. (\ref{zetaspin}) we have (to first order in $\hat s$)
\beq
y=u\left(1-\frac32{\hat s}u^{3/2}\right)^{2/3}
=\frac{u}{\left(1+\frac32{\hat s}u^{3/2}\right)^{2/3}}
\,,
\eeq
with inverse  
\beq
u=y\left(1+\frac32{\hat s}y^{3/2}\right)^{2/3}
=\frac{y}{\left(1-\frac32{\hat s}y^{3/2}\right)^{2/3}}\,.
\eeq

Finally, the conserved quantities (\ref{totalenergy}), in terms of the original (inverse) radial variable $u$ and in terms of the (invariant) frequency variable $y$ read (to first order in $\hat s$)
\begin{eqnarray}
\label{EJspin}
\frac{\bar E}{\mu}&=&\frac{1}{\sqrt{1-3u}} \left[1-2u - {\hat s}  \frac{u^{5/2}}{2(1-3u)}\right]\nonumber\\
&=&\frac{1-2y}{\sqrt{1-3y}}-\hat s \frac{y^{5/2}}{\sqrt{1-3y}}
\,,\nonumber\\
\frac{\bar J}{\mu M}&=&\frac{1}{\sqrt{u(1-3u)}} \left[1+{\hat s}\sqrt{u}\frac{1-2u}{1-3u}\left(1-\frac92u\right)\right]\nonumber\\
&=& \frac{1}{\sqrt{y(1-3y)}}+\hat s \frac{1-4y}{\sqrt{1-3y}}
\,.
\end{eqnarray}

\section{Detweiler's redshift invariant $z_1$ for a spinning particle}

The aim of the present paper is to compute  Detweiler's redshift invariant associated with a spinning particle to first order in spin, i.e., the linear-in-mass-ratio perturbation in the time component of the particle's 4-velocity to first order in both parameters $q\equiv\mu/M\ll1$ and $\hat s\ll1$.
We now consider a particle moving (according to the MPD equations) along an accelerated circular orbit but in a perturbed Schwarzschild spacetime (see Appendix B).

Let $g^{\rm R}_{\alpha\beta} = \bar g_{\alpha\beta} +qh^{\rm R}_{\alpha\beta}$ be the regularized perturbed metric (in the Detweiler-Whiting sense), where $h^{\rm R}_{\alpha\beta}$ is the regularized metric perturbation sourced by the spinning particle, which can be written as a sum of non-spinning and spinning parts, namely
\beq
h^R_{\alpha\beta}=h^{(0)}_{\alpha\beta}+\hat s h^{(\hat s)}_{\alpha\beta}\,.
\eeq 
The (perturbed) particle $4$-velocity is given by
\beq
U=u^t (\partial_t +\Omega \partial_\phi)=u^t k\,,\qquad k=\partial_t +\Omega \partial_\phi\,.
\eeq
We wish to find an expression for the gauge invariant redshift $z_1 \equiv1/u^t$. The unit normalization of the 4-velocity in the perturbed spacetime gives the condition
\begin{eqnarray}
\label{utpert}
-(u^t)^{-2}&=& \bar g_{tt} +\bar g_{\phi\phi}\Omega^2 + q h^R_{kk}\nonumber\\
&=& -\left( 1-\frac{2M}{r} \right)+r^2 \Omega^2 + q h_{kk}\,,
\end{eqnarray}
where (hereafter, we remove the label R for simplicity)
 \begin{eqnarray}
 h_{kk}&=&h_{kk}(y)=h_{\alpha\beta}k^\alpha k^\beta |_{u=y+\hat s y^{5/2}} \nonumber\\
&=&h_{kk\,(0)}(y)+{\hat s}h_{kk\,\hat s}(y)
 \end{eqnarray}  
The redshift invariant thus reads 
\begin{eqnarray} \label{z1u}
z_1(y)&=&\frac1{u^t(y)}
=\left(1-2u-\frac{y^3}{u^2}-q h_{kk}(y)\right)^{1/2}\,.
\end{eqnarray}
However, the right-handside (rhs) of this equation still contains the gauge dependent radius $u=M/r$, which must be expressed 
in terms of the gauge invariant variable $y$. 
The perturbed relation between the variables $u$ and $y$ is now given by
\beq
\label{rel_u_y}
u=\frac{y}{\left(1-\frac32{\hat s}y^{3/2}\right)^{2/3}}+q f(y)\,, 
\eeq
as a consequence of the MPD equations in the perturbed spacetime (see Appendix B),
where
\begin{align}
f(y)&=f_0(y)+\hat s f_{\hat s}(y)\,, \\
f_0(y)&=\frac1{6y}M[\partial_r h_{kk\,(0)}]^{\rm R}(y)\,,
\end{align}
and $ f_{\hat s}(y)$ will be specified in Appendix B (see, e.g., Eq. (28) of Ref. \cite{Detweiler:2008ft} for 
the derivation of $f_0(y)$).

Substituting the relation \eqref{rel_u_y} and expanding to first order in both $q$ and $\hat s$ we get
\begin{eqnarray}
\label{z1yfin}
z_1(y)&=&\sqrt{1-3y} -\frac{q}{2\sqrt{1-3y}}\left[h_{kk\,(0)}(y)\right. \nonumber\\
&&\left.+{\hat s}h_{kk\,\hat s}(y)+6\hat s y^{3/2}f_0(y)\right]\nonumber\\
&\equiv&z_1^{(0)}(y)+q \left(z_1^{(1)\hat s^0}(y)+\hat s z_1^{(1)\hat s^1}(y)\right)\,,
\end{eqnarray}
where the explicit forms of the spin-independent, and spin-linear, 1SF contributions to $z_1(y)$ (defined in
the last line) are respectively given by
\beq \label{z1yfin1}
z_1^{(1)\hat s^0}(y) = -\frac{1}{2\sqrt{1-3y}}h_{kk\,(0)}(y) \,,
\eeq
and
\beq
\label{z1yfin2}
z_1^{(1)\hat s^1}(y)=-\frac{1}{2\sqrt{1-3y}}\left[h_{kk\,\hat s}(y)+M y^{1/2}\partial_r h_{kk\,(0)}(y)\right]\,.
\eeq

Two things should be noted. First,  the spin-linear contribution $f_{\hat s}(y)$ to the $O(q)$ term $q f(y)$ in
the $ u \leftrightarrow y$ functional link \eqref{rel_u_y} has dropped out of the final results. [This follows from
the usual fact that the unperturbed value of the rhs of Eq. \eqref{z1u} is extremal with respect to $u$ (a consequence of
the geodesic character of non-spinning circular orbits).] We therefore,  do not need to explicitly compute $f_{\hat s}(y)$ (for completeness we provide, however, its formal expression in terms of regularized metric components and their derivatives in Appendix B). Second, when considering the spin-linear 1SF contribution  $z_1^{(1)\hat s^1}(y)$ 
to $z_1(y)$, there appears, besides the naively expected $h_{kk\,\hat s}(y)$ contribution, an extra term proportional
to $\partial_r h_{kk\,(0)}$. This extra term is needed to ensure the gauge-invariance of $z_1^{(1)\hat s^1}(y)$,
and its origin is the backside of what we just explained concerning the disappearance of $f_{\hat s}(y)$ in $z_1(y)$.
Indeed, as a spinning particle no longer follows a geodesic, the previous cancellation no longer (fully) operates,
and this gives rise to the last contribution in Eq. \eqref{z1yfin2}.

In the following, we shall focus on the new, spin-linear redshift contribution Eq. \eqref{z1yfin2}, and on
the computation of its regularized value
\begin{eqnarray}
\label{z1yfin2R}
z_1^{(1)\hat s^1}(y)&=&-\frac{1}{2\sqrt{1-3y}}\left[[h_{kk\,\hat s}]^R(y)\right.\nonumber\\
&& \left. +M y^{1/2}[\partial_r h_{kk\,(0)}]^R(y)\right]\,.
\end{eqnarray}
Its determination requires the two separate GSF computations: $h_{kk\,\hat s}(y)$ and $\partial_r h_{kk\,(0)}$.
 The term involving $\partial_r h_{kk\,(0)}$ comes from the non-spinning sector, which has been discussed by the authors in previous works \cite{Bini:2015bla}. Thus for the next sections we will focus on the computation of $h_{kk\,\hat s}(y)$
 (and of its regularization).

\section{Spin-dependence of the metric perturbation and $h_{kk}$}

All of our results will be computed both in the Regge-Wheeler-Zerilli and radiation-gauge frameworks.  The details of the RWZ procedure are given in Appendix C, the ultimate outcome of which are the spherical harmonic $\ell$ modes, $h_{kk}^\ell$, of $h_{kk}$. The details of the radiation-gauge metric reconstruction will be given in a future work by some of the authors \cite{Kavanaghetalinprep}. The outcome there are the tensor harmonic modes of the full metric perturbation, from which $h_{kk}^\ell$ is easily computed. In both calculations, the main difference  with the non-spinning case lies in the stress-energy tensor, which we review next.

\subsection{The energy-momentum tensor associated with the spinning particle}

The energy momentum tensor of the spinning particle is given by
\beq
T^{\alpha\beta}=T_\mu^{\alpha\beta}+T_s^{\alpha\beta}\,,
\eeq
where
\begin{eqnarray}
T_\mu^{\alpha\beta}&=&\mu\int d\tau \frac{1}{\sqrt{-g}}U^\alpha U^\beta \delta^4\,,\nonumber\\
T_s^{\alpha\beta}&=&-\int d\tau \nabla_\gamma \left[\frac{1}{\sqrt{-g}}S^{\gamma (\alpha} U^{\beta )}\delta^4\right]\,,
\end{eqnarray}
with 
\beq
S^{\alpha\beta}={\hat s} \mu M [E_1\wedge E_3]^{\alpha\beta}\,.
\eeq
Here $\delta^4$ denotes the 4-dimensional delta function centered on the particle's world line, i.e.,
\begin{eqnarray}
\delta^4&\equiv& \delta^4(x^\alpha -x^\alpha(\tau))\nonumber\\
&=&\delta (t-u^t \tau)\delta(r-r_0)\delta (\theta-\pi/2)\delta(\phi-\Omega t)\nonumber\\
&\equiv & \delta (t-u^t \tau)\delta^3\,.
\end{eqnarray}
We find then 
\begin{eqnarray}
T_\mu^{\alpha\beta}&=&\frac{\mu}{r_0^2 u^t} U^\alpha U^\beta \delta^3\,, \nonumber\\
T_s^{\alpha\beta}&=&- \nabla_\gamma \left[\frac{1}{u^t}\frac{S^{\gamma (\alpha}  U^{\beta )} }{r_0^2} \delta^3 \right]\,,
\end{eqnarray}
so that the total energy-momentum tensor finally reads
\begin{eqnarray}
\label{Tmunufin}
T_{\alpha\beta}&=&\mu \left[X^{(0)}_{\alpha\beta}+{\hat s}M X^{(s)}_{\alpha\beta}\right]\delta^{3}\nonumber\\
&&+{\hat s}\mu M\left[Y^{(s)}_{\alpha\beta}\delta_{r}^{3}+Z^{(s)}_{\alpha\beta}\delta_{\phi}^{3}\right]
\,,
\end{eqnarray} 
where
\begin{eqnarray}
\label{deltas}
\delta_{r}^{3}&=&\delta{}'(r-r_0)\delta (\theta-\pi/2) \delta (\phi-\Omega t) \,,\nonumber\\  
\delta_{\phi}^{3}&=&\delta(r-r_0)\delta (\theta-\pi/2)\delta{}'(\phi-\Omega t)\,.
\end{eqnarray}
The various contributions are given by 
\beq
X^{(0)}_{\alpha\beta} =\frac{\mu u^t}{r_0^2}\left(
\begin{array}{cccc}
f_0^2& 0& 0&  -r_0^2f_0\Omega \cr
0& 0& 0& 0\cr
0& 0& 0& 0 \cr
-r_0^2f_0\Omega& 0& 0& r_0^4\Omega^2
\end{array}
\right)\,,
\eeq
\begin{widetext}
\beq
X^{(s)}_{\alpha\beta} =\Gamma_K\left(
\begin{array}{cccc}
\displaystyle\frac{f_0\Omega_K}{r_0^3}(-2r_0+7M)& 0& 0&  -\displaystyle\frac{M^2}{r_0^4} \cr
0&-\displaystyle\frac{\Omega_K}{f_0r_0^2\Gamma_K^2}& 0& 0\cr
0& 0& 0& 0 \cr
-\displaystyle\frac{M^2}{r_0^4}& 0& 0& f_0\Omega_K
\end{array}
\right)\,,
\eeq
\beq
Y^{(s)}_{\alpha\beta} =\Gamma_Kf_0\left(
\begin{array}{cccc}
-\displaystyle\frac{f_0\Omega_K}{r_0}& 0& 0&  \displaystyle\frac{r_0-M}{2r_0^2} \cr
0& 0& 0& 0\cr
0& 0& 0& 0 \cr
\displaystyle\frac{r_0-M}{2r_0^2} & 0& 0& -r_0\Omega_K
\end{array}
\right)\,, \qquad
Z^{(s)}_{\alpha\beta} =\frac{1}{2r_0^3\Gamma_K}\left(
\begin{array}{cccc}
0& -1& 0& 0\cr
-1& 0& 0& \displaystyle\frac{r_0^2\Omega_K}{f_0}\cr
0& 0& 0& 0 \cr
0& \displaystyle\frac{r_0^2\Omega_K}{f_0}& 0& 0
\end{array}
\right)\,,
\eeq
\end{widetext}
where terms of the form $f(r)\delta'(r-r_0)$ have been replaced by $f(r_0)\delta'(r-r_0)-f'(r_0)\delta(r-r_0)$.
In the spin contributions (and only in them), the orbital frequency $\Omega$ has been replaced (consistently with the 
linear in spin approximation) by $\Omega_K$. Here, the
 subscript $K$ denotes the corresponding Keplerian (geodesic) values of $u^t$ and $\Omega$ corresponding to a spinless particle, i.e., 
\beq
\label{nuKdef}
\Gamma_K=\frac{1}{\sqrt{1-\frac{3M}{r_0}}}\,,\qquad 
\Omega_K=\sqrt{\frac{M}{r_0^3}}\,,
\eeq
and $f_0=f(r_0)$.

Decomposing  the energy momentum tensor \eqref{Tmunufin} on the tensor harmonic basis and Fourier transforming (in time), then leads to the source terms $S_{lm\omega}^{\rm (even/odd)}(r)$ entering the Regge-Wheeler equation governing both even-type and odd-type perturbations.

\section{GSF-PN expansion of the spinning redshift}

The bulk of this section will be devoted to the main new result of this paper, a post-Newtonian expansion of the spin dependence of $h_{kk}(y)$, the metric perturbation twice contracted with the helical Killing vector, considered as a
function of the orbital-frequency parameter $y$. 

\subsection{Retarded and Regularized $h_{kk}$}

The outcome of the post-Newtonian RWZ and radiation-gauge approaches are the $\ell$-modes of the retarded value of $h_{kk}$, labeled $h_{kk}^\ell$ for $\ell\geq 2$. Specifically, as detailed in previous works, we obtain explicit PN series for certain low values of $\ell=2,\ldots,6$, and generic-form solutions as a function of $\ell$ that are valid for all values $\ell\geq6$. These, when supplemented by the low multipoles $\ell=0,1$ (discussed below), yield the full retarded solution 
\beq
h_{kk}=\sum_{\ell=0}^\infty h_{kk}^\ell\,. 
\eeq
This sum is found to diverge due to the singular nature of the (spinning point particle) source. Though we are discussing here a quantity
which does not involve derivatives of the metric, we would {\it a priori} expect the large-$\ell$ behavior of the modes 
to take the form
\begin{align} \label{largel}
	h_{kk}^\ell\sim \pm A_\infty(2\ell+1)+B_\infty+\mathcal{O}(\ell^{-2})\,,
\end{align}
because the source of $h_{\mu \nu}$ contains (for a spinning particle) the derivative of a $\delta$ function. Here,
 the sign of the $A$-term depends, as usual, whether the involved radial limit is taken from above or from below.
 Our explicit computations found that the value of the $A_\infty$-coefficient  happened to be zero both 
 in Regge-Wheeler gauge, and in radiation gauge.

The expected large-$\ell$ behavior \eqref{largel} suggests to evaluate the  regularized value $h_{kk}^{\rm R}$ of $h_{kk}$
by working with the average between the two radial limits, namely
\beq \label{reghkk}
h_{kk}^{\rm R}=\sum_\ell\left[\frac12(h_{kk\,(+)}^\ell+h_{kk\,(-)}^\ell)-B_\infty\right]\, ,
\eeq
where $h_{kk\,(\pm)}^l$ are the left and right contributions.

Here, we have reasoned as if we were working in a gauge which is regularly related to the Lorenz gauge, and as if we were using a decomposition in scalar spherical harmonics (in which cases the results \eqref{largel} and \eqref{reghkk}
would follow from well-known GSF results). Actually, there are two subtleties: (i) the gauges we use are not
regularly related to the Lorenz gauge, and (ii) we use a decomposition in tensorial spherical harmonics. Concerning
the first point, we are relying on the fact that we are computing a gauge-invariant
quantity, which we could have, in principle, computed in a Lorenz gauge, and concerning the second point, 
we are relying on the fact that working with the averaged value of $h_{kk}$ effectively reduces the problem to the regularization of a field having a simpler singularity structure, which is regularized by an $\ell$-independent
$B_\infty$-type subtraction. [For a recent discussion of these subtleties in the case of the spin-precession
invariant, see, e.g., Sec.~III E of Ref. \cite{Kavanagh:2017wot}, and references therein.] 
Pending a rigorous formal justification\footnote{In addition, having analytically derived regularization parameters
would be numerically useful by providing explicit strong-field subtraction terms.} of our procedure, we wish to note here that we shall provide two different
checks of our regularization procedure: (1) our two independent calculations in two different gauges have
yielded the same final results; and (2) the first three\footnote{We count here the term of order $y^{5/2}$ that
cancels out in the final result, after appearing in intermediate calculations.} terms of our final results agree with independently calculated results in the post-Newtonian literature.

As a sample we give the form of the generic-$\ell$ results from the RWZ approach for some low-PN orders. Splitting the two contributions due to mass and spin, i.e., 
\beq
h_{kk\,(\pm)}^\ell(y)=h_{kk\,(0)\,(\pm)}^\ell(y)+{\hat s}\,h_{kk\,\hat s\,(\pm)}^\ell(y)\,,
\eeq
for $\ell\geq2$ we find
\begin{eqnarray}
&&h_{kk\,(0)\,(+)}^\ell=h_{kk\,(0)\,(-)}^\ell=2y-\frac{(26\ell^2+26\ell+3)}{(2\ell-1)(2\ell+3)}y^2\nonumber\\
&&+3\frac{(6\ell^6+18\ell^5+98\ell^4+166\ell^3+761\ell^2+681\ell-960)}{4(2\ell-3)(2\ell-1)\ell(\ell+1)(2\ell+3)(2\ell+5)}y^3\nonumber\\
&&+O(y^4)\,,
\end{eqnarray}
and
\begin{eqnarray}
&&h_{kk\,\hat s\,(+)}^\ell=h_{kk\,\hat s\,(-)}^\ell=3\frac{(\ell^2+\ell+3)}{(2\ell-1)(2\ell+3)}y^{7/2}\nonumber\\
&&-3\frac{(10\ell^6+30\ell^5+21\ell^4-8\ell^3+414\ell^2+423\ell+720)}{2(2\ell-3)(2\ell-1)\ell(\ell+1)(2\ell+3)(2\ell+5)}y^{9/2}\nonumber\\
&&+O(y^{11/2})\,.
\end{eqnarray}
Our $B_\infty$ is given by expanding these about $\ell=\infty$, order by order in the PN expansion.

\subsubsection{Low multipoles $\ell=0,1$}

When the source is a non-spinning point particle,  Zerilli \cite{Zerilli:1971wd} has shown long ago how to compute
both the exterior and the interior metric perturbations by explicitly solving the inhomogeneous RWZ field equations.
[See also Ref. \cite{Shah:2012gu} for the corresponding exterior metric computation in the case of a Kerr perturbation.]
Here, we have generalized the work of Zerilli to the case of a spinning particle, and we have determined both the
exterior and the interior metric perturbations in a RW-like gauge.  Our  derivation, and our explicit results, are
 given in Appendix A. Let us highlight here the most important aspects of our results. 
 
First,  the relevant components of the exterior metric perturbation are found (as expected) to come from the 
additional (conserved)  energy and angular momentum contribution of the spinning particle, namely
\beq
h_{tt\,(+)}^{\ell=0,1}=\frac{2\delta M}{r}\,, \qquad
h_{t\phi\,(+)}^{\ell=0,1}=-\frac{2\delta J}{r}\,,
\eeq
where $\delta M \equiv \bar E $ and $\delta J \equiv \bar J $ are given by the Killing energy and angular momentum \eqref{EJspin} of the spinning particle, respectively (see Appendix A for details).

The unsubtracted contribution to $h_{kk\,(+)}$ at the particle's location due to low multipoles is then given by
\begin{eqnarray}
h_{kk\,(+)}^{\ell=0,1}&=&h_{tt\,(+)}^{\ell=0,1}+2\Omega h_{t\phi\,(+)}^{\ell=0,1}
=\frac{2\delta M}{r_0}-\frac{4\Omega}{r_0}\delta J\nonumber\\
&=&\frac{2u(1-4u)}{\sqrt{1-3u}}-{\hat s}\frac{u^{5/2}(4-31u+54u^2)}{(1-3u)^{3/2}}\nonumber\\
&=&\frac{2y(1-4y)}{\sqrt{1-3y}}-2{\hat s}y^{5/2}\sqrt{1-3y}
\,,
\end{eqnarray}
to first order in $\hat s$.
To determine the needed left-right average $\frac12 \left( h_{kk\,(+)}^{\ell=0,1} +  h_{kk\,(-)}^{\ell=0,1}\right)$,
we further need to determine the interior metric perturbation. This is done in Appendix A. Let us cite here the
corresponding  jump of the metric components across $r=r_0$.
The RWZ equations for $\ell=0$ and $\ell=1$-odd are found to imply 
\beq
[h_{kk}^{\ell=0,1}]=h_{kk\,(+)}^{\ell=0,1}-h_{kk\,(-)}^{\ell=0,1}
=-2\hat s\frac{y^{5/2}}{\sqrt{1-3y}}\,,
\eeq
whereas $\ell=1$-even is a gauge mode having no contribution to $h_{kk}$ (see Appendix A for details).

The final result is then
\beq
\label{hkkleq01}
\frac12 \left( h_{kk\,(+)}^{\ell=0,1} +  h_{kk\,(-)}^{\ell=0,1}\right)=\frac{2y(1-4y)}{\sqrt{1-3y}}-\hat s\frac{y^{5/2}(1-6y)}{\sqrt{1-3y}}\,,
\eeq
which should still be subtracted as for the other $\ell\ge 2$ multipoles.

\subsection{Final results for $h_{kk}$ in the two gauges}

The subtraction term in the RW gauge is found to be
\beq
B_\infty=B_{(0)}+{\hat s}B_{\hat s}\,,
\eeq
with 
\begin{eqnarray}
B_{(0)}&=&2y-\frac{13}{2}y^2+\frac{9}{32}y^3+\frac{83}{128}y^4
+\frac{12361}{8192}y^5\nonumber\\
&&+\frac{116163}{32768}y^6 +\frac{4409649}{524288}y^7+\frac{42267411}{2097152}y^8\nonumber\\
&&+\frac{26189878473}{536870912}y^9+O(y^{10})
\,,
\end{eqnarray}
and
\begin{eqnarray}
B_{\hat s}&=&\frac{3}{4}y^{7/2}-\frac{15}{16}y^{9/2}-\frac{915}{512}y^{11/2}-\frac{6885}{2048}y^{13/2}\nonumber\\
&&-\frac{406755}{65536}y^{15/2}-\frac{2921697}{262144}y^{17/2}\nonumber\\
&&-\frac{321445935}{16777216}y^{19/2}+O(y^{21/2})\,.
\end{eqnarray}

After regularization, using the PN solution for $\ell>6$ and the MST solutions for $\ell=2,3,4,5,6$ (see, e.g., Ref. \cite{Bini:2013rfa} for details) and adding the low multipole contribution \eqref{hkkleq01}, we finally get
\beq
h_{kk}^{\rm R}=h_{kk\,(0)}^{\rm R}+{\hat s}h_{kk\,\hat s}^{\rm R}\,,
\eeq
with
\begin{widetext}
\begin{eqnarray}
h_{kk\,\hat s}^{\rm R}&=&
-y^{5/2}+\frac{9}{2}y^{7/2}-\frac{3}{8}y^{9/2}
+\left(\frac{189}{16}+\frac{41}{32}\pi^2\right)y^{11/2}\nonumber\\
&&
+\left(\frac{112535}{384}+\frac{672}{5}\gamma+\frac{4064}{15}\ln(2)+\frac{336}{5}\ln(y)-\frac{5533}{128}\pi^2\right)y^{13/2}\nonumber\\
&&
+\left(\frac{222734969}{44800}-\frac{552721}{1024}\pi^2-\frac{10152}{35}\gamma+\frac{2187}{7}\ln(3)-\frac{2728}{3}\ln(2)-\frac{5076}{35}\ln(y)\right)y^{15/2}\nonumber\\
&&
+\frac{217424}{1575}\pi y^8\nonumber\\
&&
+\left(-\frac{72245337401}{14515200}-\frac{439984}{567}\ln(2)-\frac{837392}{405}\gamma-\frac{181521}{70}\ln(3)-\frac{418696}{405}\ln(y)\right.\nonumber\\
&&
\left.
+\frac{182650175}{221184}\pi^2+\frac{1052215}{65536}\pi^4\right)y^{17/2}\nonumber\\
&&
-\frac{3628927}{11025}\pi y^9\nonumber\\
&&
+\left(-\frac{4331056512890369}{26078976000}+\frac{48828125}{28512}\ln(5)+\frac{72281079}{12320}\ln(3)+\frac{4548127007}{363825}\gamma+\frac{4548127007}{727650}\ln(y)\right.\nonumber\\
&&
\left.
+\frac{7042553383}{779625}\ln(2)+\frac{157132768967}{167772160}\pi^4+\frac{1182637137191}{165150720}\pi^2-\frac{2335168}{525}\ln(2)^2-\frac{29104}{105}\ln(y)^2\right.\nonumber\\
&&
\left.
-\frac{116416}{105}\gamma^2+2176\zeta(3)-\frac{116416}{105}\gamma\ln(y)-\frac{499904}{225}\ln(2)\ln(y)-\frac{999808}{225}\gamma\ln(2)\right)y^{19/2}\nonumber\\
&&+O(y^{10})\,.
\end{eqnarray}
\end{widetext}

When doing the computation in the radiation gauge, we find that the subtraction terms are identical. The 
regularized value of $h_{kk\,\hat{s}}$ is, however, different. Let us give here the difference $\Delta h_{kk\,\hat{s}}^R=h_{kk\,\hat{s}}^{R,\text{RG}}-h_{kk\,\hat{s}}^R$, where RG labels the radiation gauge result:
\begin{align}
	\Delta h_{kk\,\hat{s}}^R&=\left(33-4 \pi ^2\right) y^{11/2}+\left(-\frac{1317}{50}+2 \pi ^2\right) y^{13/2} \nonumber \\
	&+\left(\frac{181883}{9800}-\frac{3 \pi ^2}{2}\right)
   y^{15/2}-\frac{128 \pi  y^8}{5} \nonumber \\ 
   &+\left(\frac{2287038017}{952560}-\frac{9011 \pi ^2}{36}+\frac{16 \pi ^4}{15}\right) y^{17/2}\nonumber \\ 
   &+O\left(y^9\right)\,.
\end{align}
This difference is, however, a gauge effect that will disappear when computing the gauge-invariant quantity $z_1^{(1)\hat s^1}$.

\subsection{Final results for $\partial_r h_{kk\,(0)}$ in the two gauges}

The computation of $\partial_r h_{kk\,(0)}$ proceeds exactly as in the case of $h_{kk \hat s}$.
We then skip all unnecessary details and display only the final result, which in the Regge-Wheeler gauge is:
\begin{widetext}
\begin{eqnarray}
M[\partial_r h_{kk\,(0)}]^{\rm R}(y)&=&
 y^2-\frac{13}{2}y^3+\frac{75}{8}y^4+\left(-\frac{41}{32}\pi^2-\frac{57}{16}\right)y^5\nonumber\\
&&
+
\left(\frac{191101}{1920}-\frac{512}{5}\gamma-\frac{1024}{5}\ln(2)-\frac{256}{5}\ln(y)+\frac{1661}{512}\pi^2\right)y^6
\nonumber\\
&&
+\left(\frac{26793971}{44800}-\frac{1458}{7}\ln(3)+\frac{5168}{7}\ln(2)+\frac{1840}{7}\gamma+\frac{920}{7}\ln(y)-\frac{39495}{1024}\pi^2\right)y^7\nonumber\\
&&
-\frac{54784}{525}\pi y^{15/2}\nonumber\\
&&
+\left(\frac{159402781889}{14515200}-\frac{2800873}{262144}\pi^4-\frac{2367261307}{1769472}\pi^2+\frac{3611672}{2835}\gamma+\frac{1805836}{2835}\ln(y)\right.\nonumber\\
&&\left.
+\frac{1064408}{2835}\ln(2)+1701\ln(3)\right)y^8\nonumber\\
&&
+\frac{353898}{1225}\pi y^{17/2}\nonumber\\
&&
+\left(\frac{438272}{525}\gamma\ln(y)-\frac{9765625}{9504}\ln(5)-\frac{12471233664763}{2477260800}\pi^2+\frac{245032783}{16777216}\pi^4-\frac{51161269282}{5457375}\gamma\right.\nonumber\\
&&\left.
-\frac{47957923714}{5457375}\ln(2)-\frac{25580634641}{5457375}\ln(y)-\frac{9225009}{2464}\ln(3)+\frac{438272}{525}\gamma^2+\frac{109568}{525}\ln(y)^2\right.\nonumber\\
&&\left.
+\frac{1753088}{525}\ln(2)^2-\frac{8192}{5}\zeta(3)+\frac{876544}{525}\ln(2)\ln(y)+\frac{1753088}{525}\gamma\ln(2)+\frac{20855431768697683}{391184640000}\right)y^9\nonumber\\
&&
+\frac{3923438969}{3274425}\pi y^{19/2}
+O(y^{10})\,.
\end{eqnarray}
The subtraction term in this case turns out to be (in both gauges)
\beq
B_\infty= -y^2+\frac{11}{4}y^3+\frac{27}{64}y^4+\frac{199}{256}y^5+\frac{22783}{16384}y^6
+\frac{155475}{65536}y^7+\frac{3899547}{1048576}y^8+\frac{20318463}{4194304}y^9
+O(y^{10})\,.
\eeq
Again, defining the difference with the radiation gauge as $\Delta \partial_r h_{kk\,(0)}^R=\partial_r h_{kk\,(0)}^{R,\text{RG}}-\partial_r h_{kk\,(0)}^R$, we find
\begin{align}
	M \Delta \partial_r h_{kk\,(0)}^R&=-\left(33-4 \pi ^2\right) y^{5}-\left(-\frac{1317}{50}+2 \pi ^2\right) y^{6} -\left(\frac{181883}{9800}-\frac{3 \pi ^2}{2}\right)
   y^{7}+\frac{128 \pi  y^{15/2}}{5} \nonumber \\ 
   &-\left(\frac{2287038017}{952560}-\frac{9011 \pi ^2}{36}+\frac{16 \pi ^4}{15}\right) y^{8}+O\left(y^{17/2}\right).
\end{align}
Importantly, we note that this is exactly $-y^{-1/2} \Delta h_{kk\,\hat{s}}^R$. In view of Eq. \eqref{z1yfin2}
this will ensure the gauge-independence of our final result for $z_1^{(1)\hat s^1}$.

\subsection{Final result for $z_1^{(1)\hat s^1}$}

The linear in spin correction to Detweiler's gauge-invariant redshift function finally reads 
\begin{eqnarray}
	\label{Eq:z1sgsf}
z_1^{(1)\hat s^1}(y)&=&
y^{7/2}-3y^{9/2}-\frac{15}{2}y^{11/2}+
\left(-\frac{6277}{30}+\frac{20471}{1024}\pi^2-16\gamma-\frac{496}{15}\ln(2)-8\ln(y)\right)y^{13/2}\nonumber\\
&&
+\left(\frac{653629}{2048}\pi^2-\frac{87055}{28}-\frac{729}{14}\ln(3)+\frac{3772}{105}\ln(2)-\frac{52}{5}\gamma-\frac{26}{5}\ln(y)\right)y^{15/2}\nonumber\\
&&
-\frac{26536}{1575}\pi y^8\nonumber\\
&&
+\left(-\frac{149628163}{18900}+\frac{4556}{21}\ln(2)+\frac{7628}{21}\gamma+\frac{12879}{35}\ln(3)+\frac{3814}{21}\ln(y)+\frac{297761947}{393216}\pi^2-\frac{1407987}{524288}\pi^4\right)y^{17/2}\nonumber\\
&&
-\frac{113411}{22050}\pi y^9\nonumber\\
&&
+\left(-\frac{74909462}{70875}\gamma+\frac{340681718}{1819125}\ln(2)-\frac{199989}{352}\ln(3)-\frac{1344}{5}\zeta(3)-\frac{9765625}{28512}\ln(5)+\frac{164673979457}{353894400}\pi^2\right.\nonumber\\
&&\left.
-\frac{160934764317}{335544320}\pi^4+\frac{3424}{25}\gamma^2+\frac{58208}{105}\ln(2)^2+\frac{869696}{1575}\gamma\ln(2)-\frac{37454731}{70875}\ln(y)+\frac{3424}{25}\gamma\ln(y)\right.\nonumber\\
&&\left.
+\frac{434848}{1575}\ln(2)\ln(y)+\frac{856}{25}\ln(y)^2+\frac{403109158099}{9922500}\right)y^{19/2}
+O(y^{10})\,.
\end{eqnarray}
\end{widetext} 
This is the main result of the present paper. Importantly, as we already said, this final result is (as expected for a gauge-invariant quantity) identical between the two gauges we have worked in.

\begin{figure}
\includegraphics[scale=0.35]{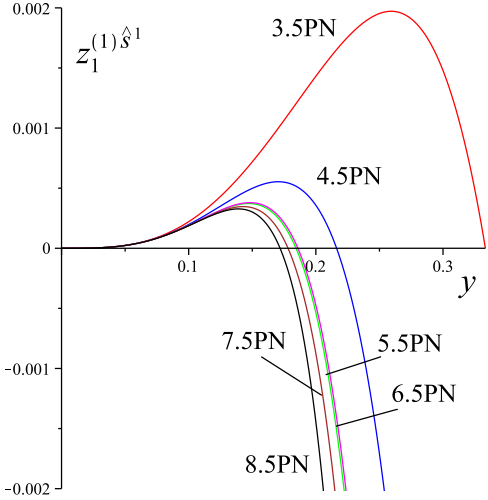}
\caption{\label{fig:1} Some of the various PN approximants to the linear-in-spin 1SF contribution to the redshift function $z_1(y)$ for a spinning particle moving along a circular orbit in a Schwarzschild spacetime.}
\end{figure}

We show in Fig. \ref{fig:1} the behavior of the various PN approximants to $z_1^{(1)\hat s^1}(y)$, which becomes more and more negative as the light-ring is approached, thereby suggesting a negative power-law divergence there.


\subsection{Comparison with PN results}

In PN theory the linear-in-spin part of the Hamiltonian (and therefore, using Ref. \cite{Blanchet:2012at}, the
corresponding linear-in-spin part, $z_1^{(1)\hat s^1}$, of the redshift $z_1= \partial H/\partial m_1$), 
is known up to the next-to-next-to-leading order \cite{Levi:2015uxa}.
Using the results of \cite{Levi:2015uxa}, we have computed $z_1^{(1)\hat s^1}$ as a function of 
$x\equiv ((M+\mu) \Omega)^{2/3}$, with the following result (corresponding to the 3.5PN order):
\begin{eqnarray}
z_1^{(1)\hat s^1}(x)&=&\sum_{k=2}^\infty C_{\chi_1}^{(2k+1)/2}(\nu; \ln x)\chi_1 x^{(2k+1)/2}\,,\nonumber\\
\end{eqnarray}
where the coefficients are given by
\begin{eqnarray}
C_{\chi_1}^{5/2} &=& \frac13\nu\Delta -\frac13 \nu+\frac23 \nu^2\,, \nonumber\\
C_{\chi_1}^{7/2} &=&  -\frac12 \nu\Delta +\frac{19}{18}\nu^2-\frac{19}{18}\nu^2\Delta -\frac19\nu^3+\frac12\nu  \,, \nonumber\\
C_{\chi_1}^{9/2} &=&   -\frac{39}{8}\nu^2+\frac{11}{24}\nu^3\Delta +\frac{27}{8}\nu-\frac{1}{12}\nu^4-\frac{27}{8}\nu\Delta \nonumber\\
&&-\frac{161}{24}\nu^3-\frac{39}{8}\nu^2\Delta\,.
\end{eqnarray}

Here $\chi_1\equiv S_1/\mu^2$, $\nu\equiv\mu M/M_{\rm tot}^2$ and $\Delta \equiv (M-\mu)/M_{\rm tot}=\sqrt{1-4\nu}$, with $M_{\rm tot}= M+\mu$.
To convert this result into the 1SF contribution to  $z_1(y)$, we use:
$S_1=\mu M {\hat s}$, $q=\mu/M$, $\mu=M_{\rm tot}(1-\Delta)/2$, $M=M_{\rm tot}(1+\Delta)/2$, and $x=(1+q)^{2/3}y$.
The first term $C_{\chi_1}^{5/2}$  does not contribute at the first order in $q$, i.e., at the first order in SF expansion, while the last two terms yield
\beq
\label{z1_s_PN}
z_1^{(1)\hat s^1}={\hat s}y^{7/2}\left(1-3y+O(y^2)\right)q+O(q^2)\,.
\eeq
This agrees with the first two terms of \eqref{Eq:z1sgsf}, thereby providing an independent (partial) check
of our result.

\section{Concluding remarks}

The original contribution of this paper is the formulation of the generalization of Detweiler's redshift function $z_1(y)$ 
for a spinning particle on a circular orbit in Schwarzschild, and its first computation at a high PN-order (8.5PN, instead of the 
currently known 3.5PN order).
The spinning particle moves here along an accelerated orbit, deviating from a timelike circular geodesics because of the spin itself which couples to the Riemann tensor of the background. We have shown how this non-geodesic character of the orbit 
induces in the spin-linear contribution to $z_1(y)$ a (gauge-dependent) term proportional to the radial gradient of $h_{k k}$ which plays
a crucial role in ensuring the gauge-invariance of the final result. We have checked the gauge-invariance of our result
by providing a dual calculation, in two different gauges, and in verifying that the final results agree.
Our formulation opens the way to strong field numerical studies and provides a benchmark for their results. It would also be of interest to have independent investigations of the regularization procedure we use.

Another original result of this work (essential to accomplish the first result) has been the \lq\lq completion" of the perturbed metric by the explicit computation (in the Regge-Wheeler gauge) of the contribution of the non-radiative multipoles to both the interior and exterior metric generated by a spinning particle. We expect this result to
play a useful role in future applications.

Finally, using available PN results, we have checked the first terms of our final result.

\appendix

\section{Low multipoles $l=0,1$}

We give below the solutions for the non-radiative modes ($\ell=0$ and $\ell=1$ odd) needed for the completion of the full metric perturbation. Our approach is a generalization of well-known results of Zerilli \cite{Zerilli:1971wd} to the
case of a spinning particle. 
The $\ell=1$ even mode is essentially a gauge mode that describes a shift of the center of momentum of the system.
We have checked that it does not contribute to the present calculation.

\subsection{The $\ell=0$ mode}

The $\ell=0$ mode is of even parity, is independent of time and represents the perturbation in the total mass-energy of the system. This was shown by Zerilli for the case of a non-spinning test particle, and our explicit calculations below
show that this extends to the case of a spinning particle if one uses as additional contribution to the mass
of the system the conserved Killing energy $\delta M\equiv \bar E$, Eq. \eqref{totalenergy}, \eqref{EJspin}, of the spinning particle. Note that our derivation directly solves the inhomogeneous Regge-Wheeler-Zerilli equations, without
using Komar-type surface integrals.

For this mode there are two gauge degrees of freedom and one can set $H_1=0=K$.
The remaining perturbation functions $H_0$ and $H_2$ satisfy the following equations
\begin{widetext}
\begin{eqnarray}
\frac{dH_2}{dr}+\frac{H_2}{rf}&=&2\sqrt{4\pi}\mu\frac{u^t}{r_0}\left[
\delta(r-r_0)-M\Omega_K\hat s\left(\frac{r_0-M}{r_0f_0}\delta(r-r_0)+r_0\delta'(r-r_0)\right)
\right]\,,\nonumber\\
\frac{dH_0}{dr}+\frac{H_2}{rf}&=&2\sqrt{4\pi}\mu\frac{M\Omega_K}{r_0\Gamma_Kf_0}\hat s\delta(r-r_0)
\,,
\end{eqnarray}
to first order in $\hat s$, with solution 
\begin{eqnarray}
H_0&=&2\sqrt{4\pi}\mu u^t\left[
\frac{1}{r_0}(1-2{\hat s}M\Omega_K)\theta (r_0-r)+\frac{f_0}{rf}\left(1+{\hat s}\frac{M^2\Omega_K}{r_0f_0}\right)\theta (r-r_0)
\right]\,,\nonumber\\
H_2&=&2\sqrt{4\pi}\mu u^t\left[
\frac{f_0}{rf}\left(1+{\hat s}\frac{M^2\Omega_K}{r_0f_0}\right)\theta (r-r_0)-{\hat s}M\Omega_K\delta(r-r_0)
\right]\,.
\end{eqnarray}
The nonvanishing metric components to first order in $\hat s$ can  then be written (in terms of 
$\delta M\equiv \bar E$, Eq. \eqref{EJspin}) as
\begin{eqnarray}
h_{tt}&=&\frac{fH_0}{\sqrt{4\pi}}=\frac{2\mu\, \delta M}{r}\left[
\frac{rf}{r_0f_0}\left(1-\frac{2r_0-3M}{r_0f_0}M\Omega_K\hat s\right)\theta (r_0-r)+\theta (r-r_0)
\right]\,,\nonumber\\
h_{rr}&=&\frac{H_2}{\sqrt{4\pi}f}=\frac{2\mu\, \delta M}{rf^2}\theta (r-r_0)-\frac{2\mu\,\delta M}{f_0^2}M\Omega_K{\hat s}\delta(r-r_0)\,.
\end{eqnarray}

\end{widetext}

\subsection{The $\ell=1$ odd mode}

Similarly, we have explicitly shown, by solving the Regge-Wheeler-Zerilli field equations, that the $\ell=1$ 
odd mode represents the angular momentum perturbation $\delta J= \bar J$, Eq. \eqref{totalenergy}, \eqref{EJspin}, 
added by the spinning particle to the system. 

The perturbation equations for this case assume $h_1^{\rm(odd)}=0$, whereas $h_0^{{\rm(odd)}}$ is such that
\begin{widetext}
\beq
\frac{d^2h_0^{{\rm(odd)}}}{dr^2}-\frac{2h_0^{{\rm(odd)}}}{r^2}=-4\mu\sqrt{3\pi}u^t\Omega\left[
\delta(r-r_0)-M\Omega_K\hat s\left(\delta(r-r_0)+\frac{r_0}{2M}(r_0-M)\delta'(r-r_0)\right)
\right]\,,
\eeq
to first order in $\hat s$, with solution 
\beq
h_0^{\rm(odd)}=2\sqrt{\frac{4\pi}{3}}\mu u^t\Omega r_0\left[\frac{r^2}{r_0^2}\left(1-\frac12{\hat s}(r_0+M)\Omega_K\right)\theta (r_0-r)+\frac{r_0}{r}(1+{\hat s}r_0\Omega_Kf_0)\theta (r-r_0)\right]\delta_{m,0}\,,
\eeq
and only the $m=0$ mode is nonzero. 
The only nonvanishing metric component is then given (in terms of 
$\delta J \equiv \bar J$, Eq. \eqref{EJspin}) by
\beq
h_{t\phi}=-\sqrt{\frac{3}{4\pi}}h_0^{{\rm(odd)}}\sin^2\theta
=-\frac{2\mu\,\delta J}{r}\left[\frac{r^3}{r_0^3}\left(1-\frac32(r_0-M)\Omega_K{\hat s}\right)\theta (r_0-r)+\theta (r-r_0)\right]\sin^2\theta\,,
\eeq
to first order in $\hat s$.
\end{widetext}

\section{MPD equations in the perturbed spacetime}

In this section we briefly discuss the MPD equations in the perturbed spacetime to first-order in spin.
This complementary material is left here for convenience and it will be of use in future works.
Working to the first order in spin, Eqs. (\ref{papcoreqs1}) and (\ref{papcoreqs2}) reduce to
\begin{eqnarray}
\label{papcoreqs1_A}
\mu \frac{{\rm D}U^{\mu}}{d \tau} & = &
- \frac12 \, R^\mu{}_{\nu \alpha \beta} \, U^\nu \, S^{\alpha \beta} \,,
\\
\label{papcoreqs2_A}
\frac{{\rm D}S^{\mu\nu}}{d \tau} & = & 0\,,
\end{eqnarray}
where we recall that $U^{\mu} \equiv d z^\mu/d \tau$, and where we have used the property $P=\mu U+O(s^2)$ for the momentum of the particle. 

Assuming that the background metric admits the Killing vector $k=\partial_t+\Omega\partial_\phi$ and that the body's orbit is aligned with $k$, $U=u^t k$, implying
\beq
\label{Gamma_def}
-(u^t)^{-2}=k\cdot k=-f +\Omega^2 r^2 +h_{kk}\,,
\eeq
the MPD equations become
\begin{eqnarray}
\mu u^t  \nabla_k k^{\mu}=
- \frac12 \, R^\mu{}_{\nu \alpha \beta} \, k^\nu \, S^{\alpha \beta}\,,\quad 
\nabla_U S^{\mu\nu}=0\,.
\end{eqnarray}
Defining then the spin vector (orthogonal to both $U$ and $u \equiv P/\mu$ at the first order in spin) by spatial duality (see Eq. \eqref{spinvecdef})
\beq
S^\gamma 
=\frac12 u^t k_\sigma \eta^{\sigma\gamma\alpha\beta}S_{\alpha\beta}\,,\qquad S^\gamma k_\gamma=0\,,
\eeq
one finds immediately  that the spin vector is parallel-propagated along $U$, $\nabla_U S^\gamma=0$.
The equations of motion instead can be cast in the form
\begin{eqnarray}
\label{eq_moto_k}
\mu u^t  \nabla_k k^{\mu}&=&
- \frac12 ( \nabla_{\mu\beta}k_\alpha) \, S^{\alpha \beta}\nonumber\\
&=&- \frac12  (\nabla_{\mu}K_{\beta\alpha} )\, S^{\alpha \beta}\,, 
\end{eqnarray}
where  the (antisymmetric) tensor $K_{\alpha\beta}$ is given by
\beq
K_{\alpha\beta}=\nabla_\alpha k_\beta=\partial_{[\alpha}k_{\beta]}\,.
\eeq
Finally, we require that the spin vector be orthogonal to the equatorial plane, i.e.,
\beq
S=-s e_{\hat \theta}\,,\qquad
e_{\hat \theta}=\frac1r\left(1-\frac1{2r^2}h_{\theta\theta}\right)\partial_\theta\,.
\eeq
The equations of motion then imply the following solution for $\Omega$
\beq
M\Omega=M\Omega_K\left[
1-\frac32{\hat s}M\Omega_K+q(\tilde\Omega_1+{\hat s}\tilde\Omega_{1{\hat s}})
\right]\,,
\eeq
where $\Omega_K\equiv\sqrt{\frac{M}{r_0^3}}$, as defined in Eq. \eqref{nuKdef} above,
\beq
\tilde\Omega_1=
-\frac{M}{4u^2}[\partial_r h_{kk}^{(0)}]_{r=M/u}\,,
\eeq
and
\begin{widetext}
\begin{eqnarray}
\tilde\Omega_{1{\hat s}}&=&
-\frac{u^{3/2}}{4(1-2u)^2}h_{kk}^{(0)}
+\frac{(5-12u)u^{3/2}}{4}h_{rr}^{(0)}
-\frac{u^2(3-4u)(1-3u)}{2M(1-2u)^2}h_{t\phi}^{(0)}
-\frac{(1-3u)(2-5u+4u^2)u^{5/2}}{4M^2(1-2u)^2}h_{\phi\phi}^{(0)}\nonumber\\
&&
-\frac{M^2u^{-3/2}}{4}[\partial_{rr} h_{kk}^{(0)}]_{r=M/u}
-\frac{M}{4u}(1-3u)[\partial_{rr} h_{\phi k}^{(0)}]_{r=M/u}
+\frac{M}{4u}(1-3u)[\partial_{r\bar\phi} h_{rk}^{(0)}]_{r=M/u}\nonumber\\
&&
-\frac14(1-3u)[\partial_{\bar\phi} h_{rk}^{(0)}]_{r=M/u}
+\frac{Mu^{1/2}}{4(1-2u)}[\partial_r h_{kk}^{(0)}]_{r=M/u}
-\frac{M}{4u^2}[\partial_r h_{kk}^{(1)}]_{r=M/u}\nonumber\\
&&
+\frac{M}{4}(1-2u)(1-3u)u^{-1/2}[\partial_r h_{rr}^{(0)}]_{r=M/u}
+\frac{(1-3u)}{4(1-2u)}[\partial_r h_{t\phi}^{(0)}]_{r=M/u}
+\frac{(1-3u)(2-3u)u^{3/2}}{4M(1-2u)}[\partial_r h_{\phi\phi}^{(0)}]_{r=M/u}
\,.\nonumber\\
\end{eqnarray}
\end{widetext}

Introducing the dimensionless frequency parameter $y=(M\Omega)^{2/3}$ gives the relation 
\beq
y=u-{\hat s}u^{5/2}+q {\mathcal F}(u)\,, 
\eeq
where ${\mathcal F}(u)={\mathcal F}_0(u)+\hat s {\mathcal F}_{\hat s}(u)$, with 
\begin{eqnarray}
{\mathcal F}_0(u)&=&\frac23u\tilde\Omega_1(u)\,,\nonumber\\
{\mathcal F}_{\hat s}(u)&=&\frac13u^{5/2}\tilde\Omega_1(u)+\frac23u\tilde\Omega_{1{\hat s}}(u)\,.
\end{eqnarray}
This relation can be inverted to give (see Eq. \eqref{rel_u_y})
\beq
u=\frac{y}{\left(1-\frac32{\hat s}y^{3/2}\right)^{2/3}}+q f(y)\,, 
\eeq
where $f(y)=f_0(y)+\hat s f_{\hat s}(y)$, with 
\begin{eqnarray}
f_0(y)&=&-{\mathcal F}_0(y)\,,\nonumber\\
f_{\hat s}(y)&=&-{\mathcal F}_{\hat s}(y)
-\frac52y^{3/2}{\mathcal F}_0(y)-y^{5/2}{\mathcal F}'_0(y)\,.
\end{eqnarray}
Substituting then into Eq. (\ref{Gamma_def}) finally yields Eq. \eqref{z1yfin}.

\section{Metric reconstruction in the Regge-Wheeler gauge}

\subsection{Solving the RWZ equations}

The perturbation functions of both parity can be expressed in terms of a single unknown for each sector, satisfying the same Regge-Wheeler equation
\beq
\label{binieq:RW}
{\mathcal L}^{(r)}_{\rm (RW)}[R_{\ell m\omega }^{\rm (even/odd)}]=S_{\ell m\omega}^{\rm (even/odd)}(r) \,,
\eeq
where ${\mathcal L}^{(r)}_{\rm (RW)}$ denotes the RW operator
\begin{eqnarray}
\label{binieq:operator}
{\mathcal L}^{(r)}_{\rm (RW)}&=&f(r)^2 \frac{d^2}{dr^2} +\frac{2M}{r^2}f(r) \frac{d}{dr } +[\omega^2 -V_{\rm (RW)}(r)]\nonumber\\
&=& \frac{d^2}{dr_*^2}   +[\omega^2 -V_{\rm (RW)}(r)]\,,
\end{eqnarray}
with $d/dr_* = f(r) d/dr$, and the RW potential
\beq
\label{binieq:potential}
V_{\rm (RW)}(r)=f(r) \left(\frac{\ell (\ell +1)}{r^2}-\frac{6M}{r^3}  \right)\,.
\eeq
The source terms have the form
\begin{eqnarray}
\label{sourceexp}
S_{\ell m\omega}^{\rm (even/odd)}(r)&=&c_0^{\ell m\omega} \delta(r-r_0)+c_1^{\ell m\omega} \delta'(r-r_0)\nonumber\\
&+&c_2^{\ell m\omega} \delta''(r-r_0)+c_3^{\ell m\omega} \delta'''(r-r_0)\,.\nonumber\\
\end{eqnarray}
The coefficients $c_k^{\ell m\omega}$, $k=0\ldots 3$ are not depending on $r$ and have the general form
\beq
c_k^{\ell m\omega}=\tilde c_{k}^{\ell m\omega}\delta (\omega- m\Omega)\,,
\eeq
with $c_3^{\ell m\omega}\equiv0$ in the odd case.

The Green's function is expressed in terms of the two independent homogeneous solutions $X_{\ell \omega}^{{\rm in}}$ and $X_{\ell \omega}^{{\rm up}}$ of the RW operator as
\begin{eqnarray}
\label{binieq:Green}
G(r,r')&=& G_{\rm (in)}(r,r')H(r'-r)+G_{\rm (up)}(r,r')H(r-r') \nonumber \,,
\end{eqnarray}
where
\begin{eqnarray}
 G_{\rm (in)}(r,r')&=& \frac{X_{\ell \omega}^{{\rm in}}(r)X_{\ell \omega}^{{\rm up}}(r')}{W_{\ell \omega}}\,, \nonumber\\
 G_{\rm (up)}(r,r')&=& \frac{ X_{\ell \omega}^{{\rm in}}(r')X_{\ell \omega}^{{\rm up}}(r)}{W_{\ell \omega}}\,.
\end{eqnarray}
Here $W_{\ell \omega}$ denotes the (constant) Wronskian
\begin{eqnarray}
\label{binieq:wronski}
W_{\ell \omega}&=&f(r)\biggl[X_{\ell \omega}^{{\rm in}}(r)\frac{d}{dr }X_{\ell \omega}^{{\rm up}}(r )-\frac{d}{dr}X_{\ell \omega}^{{\rm in}}(r)X_{\ell \omega}^{{\rm up}}(r) \biggl]\nonumber\\
&=&{\rm const.}
\end{eqnarray}
and  $H(x)$ is the Heaviside step function.
Both even-parity and odd-parity solutions are then given by integrals over the corresponding (distributional) sources as
\beq
R^{\rm(even/odd)}_{\ell m\omega}(r)=\int d r' \, \frac{G(r,r')}{f(r')}S^{\rm(even/ odd)}_{\ell m\omega}(r')\,.
\eeq

Once the radial function is known for both parities, the perturbed metric components are then computed by Fourier anti-transforming, multiplying by the angular part and summing over $m$ (between $-\ell$ and $+\ell$), and 
then over $\ell$ (between $0$ and $ +\infty$).

\subsection{Computing $h_{kk}$}

Let us consider the quantity $h_{kk}\equiv h_{\alpha\beta}k^\alpha k^\beta$, where $k=\partial_t+\Omega\partial_\phi$.
In the RW gauge we have
\beq
h_{kk}=\sum_{\ell m}h_{kk}^{\ell m}=\sum_{\ell m}(h_{kk}^{\ell m\,\rm(even)}+h_{kk}^{\ell m\,\rm(odd)})\,,
\eeq
where the even and odd contributions (for $\ell \geq2$) are of the form
\begin{eqnarray}
h_{kk}^{\ell m\, \rm (even)}(r_0)&=& \left|Y_{\ell m}\left(\frac{\pi}{2},0\right)\right|^2 A_{\ell m}^{\rm even}(r_0) J_{\rm in}(r_0) J_{\rm up}(r_0)\,,\nonumber\\
h_{kk}^{\ell m\, \rm (odd)}(r_0) &=& \left|\partial_\theta Y_{\ell m}\left(\frac{\pi}{2},0\right)\right|^2 A_{\ell m}^{\rm odd}(r_0) \tilde J_{\rm in}(r_0) \tilde J_{\rm up}(r_0)\,,\nonumber
\end{eqnarray}
once evaluated along the world line of the particle $r=r_0$, $\theta=\pi/2$, $\phi=\Omega t$.
The coefficients $A_{\ell m}^{\rm even/odd}(r_0)$ and
\begin{eqnarray}
J_{\rm in/up}(r_0)&=&\alpha_{\rm in/up}(r_0) X^{\rm in/up}_{\ell \omega}(r_0)\nonumber\\
&&+\beta_{\rm in/up}(r_0) \left.\frac{d  X^{\rm in/up}_{\ell \omega}}{dr}\right|_{r=r_0}\,,\nonumber\\
\tilde J_{\rm in/up}(r_0)&=&\tilde\alpha_{\rm in/up}(r_0) X^{\rm in/up}_{\ell \omega}(r_0)\nonumber\\
&&+\tilde\beta_{\rm in/up}(r_0) \left.\frac{d  X^{\rm in/up}_{\ell \omega}}{dr}\right|_{r=r_0}\,,
\end{eqnarray}
all depend on $\hat s$ and are known functions of $r_0$, with $\omega=m\Omega$.

Expanding all terms to first order in ${\hat s}$ and combining the odd and even contributions leads to
\beq
h_{kk}^{\ell m}=h_{kk\,(0)}^{\ell m}+{\hat s}h_{kk\,\hat s}^{\ell m}\,.
\eeq
Performing the $m$-summation yields by definition
\beq
h_{kk}^\ell \equiv\sum_m h_{kk}^{\ell m}\,.
\eeq
One must then finally express $r_0$ in terms of $y$ and $\hat s$ to get $h_{kk}^\ell(y, \hat s)$.

\section*{Acknowledgments}

DB is grateful to Jan Steinhoff for useful exchanges of information, and for providing some of his results
in advance of publication.
DB thanks ICRANet and the italian INFN for partial support and IHES for warm hospitality at various stages during the development of the present project.

\end{document}